\documentclass[  twocolumn,aps ] { revtex4}
\usepackage{graphicx}
\usepackage{bm}
\usepackage{amsmath, mathrsfs}
\usepackage{amssymb}

\usepackage{slashed}
\usepackage{color}
\usepackage{CJK}
\usepackage{ulem}

\begin{document}
\title{  Temperature and volume dependence of   pion-pion scattering lengths  \footnote{  Supported   by the
		Fostering Program in Disciplines Possessing Novel Features for Natural Science of Sichuan University (Grant No. 2020SCUNL209). } }
\author{Qing-Wu Wang $^{1}$}
\author{Hua-Zhong Guo $^{1}$}~\email[]{Email: qw.wang@scu.edu.cn;guohuazhong@scu.edu.cn}
\affiliation{
$^1$Department of Physics,  Sichuan University,   Chengdu 610064,  China}

\begin{abstract}
 $\mathbf{Abstract:}$
The s-wave  pion-pion scattering lengths $ a_0 $ and $ a_2 $ are studied at finite temperature and  in finite spatial volume under the framework of the Nambu--Jona-Lasinio model.
With the proper time regularization,  the behavior beyond the pseudo transition temperature is   presented.  The scattering length $ a_0$ shows  singularity  near the Mott temperature and $ a_2$ is  a continuous but non-monotonic function of temperature.
We present the finite volume effect on the scattering length  and have found that $a_0$ can be negative and its  singularity  disappears at small volume size which may hint the existence of chiral phase transition as volume decreases.

$\mathbf{Keywords:}$ scattering length,  finite volume effect,   chiral phase transition
\end{abstract}

\maketitle

\section{introduction}
Dynamical chiral symmetry breaking is an important feature of Quantum Chromodynamics(QCD).
 With the rapid development in heavy ion experiment and observation in astronomy,  restoration of the chiral symmetry and the deconfinement phase transition   are expected to occur in
ultra-relativistic heavy-ion collisions   or in the interior of neutron stars  [1-4]. Many famous research institutions,  such as FAIR GSI,  NICA JINR,  and J-PARK,
have carried out experiments to study the properties of high-density matter which could help to clarify the phase structure of quark matter.

As a gauge theory at short distances,  perturbative Quantum Chromodynamics is a remarkably successful and rich theory of the strong interactions. However,  many physical phenomena in a long distance have to be dealt with non-perturbative method,  especially in handling the problems related to low-energy physics,  the non-perturbation model is of particularly great use.
 Among many effective models pion plays a crucial role with the fact that it occupies a special place in nuclear and particle physics.
   In the standard picture,  pion is explained as the (pseudo-) Nambu-Goldstone boson. It arises as a consequence of the dynamical breakdown of the chiral symmetry [5,6]. In the chiral limit,  where the masses of the two lightest quarks are turned off,  the pion mass is zero.
  When the chiral symmetry is broken spontaneously,  the
 quark condensate represents the leading order parameter.
    The Nambu-Goldstone bosons can interact if they carry momentum.
 Weinberg's low energy theorems state that pion-pion scattering  lengths  are related to the pion
 mass with $ a_0\sim7M_\pi^2/(32\pi f_\pi^2) $,  $ a_2\sim-M_\pi^2/(16\pi f_\pi^2) $.
 In the chiral
 limit,  the pion-pion
 S-wave   scattering  lengths  vanish,
and these quantities can be used as a
 sensitive probe of the chiral symmetry breaking.

  The pion-pion scattering,  as one of the most fundamental hadronic processes of QCD at the mesonic level,
 provides a direct link between the theoretical formalisms of chiral
 symmetry and experiments. Most calculations so far have been done in the infinite volume limit[7-13].
 Works on the chiral phase transition are stressed on the influence of temperature,  baryon densities and other outside parameters such as magnetic field [14-17].    But the quark-gluon plasma(QGP) system produced by
 heavy-ion collisions   experiments
 always has a finite size.
 The volume of homogeneity before freeze out for Au–Au and Pb–Pb collisions
 is between approximately $ 50 \sim250 $ fm$ ^3 $. It is  estimated
  that the volume of the smallest QGP system could be as low as (2 fm)$ ^3 $ [18],  motivated by the estimated plasma size presumably formed
  in high-energy  nucleus-nucleus collision  at  RHIC   [1 ].
   Studies show that the chiral behavior  depends on the volume size of quark matter and the chiral symmetry breaking is closely related to finite volume effects in QCD [19-22].
  Besides,  the finite size effect on the    dissociation and diffusion of chiral partners,
     on the phase structure of the NJL model in D=3 Euclidean dimensions and on the
   viscosity,  bulk viscosity,  electrical conductivity have been studied [23-25].

     Since pion meson plays an important role in low-energy physics,  it is worthwhile to clarify its behaviors as extrapolation to high temperature and small volume size.

  The chiral phase transition of a finite system depends on the choice of the boundary conditions  [26,27]. For quark fields,    the time direction is constrained by anti-period requirement,  but the choice of boundary condition in spatial direction is at large. Typical boundary conditions are anti-period boundary condition (APBC) and period boundary condition(PBC) which are used to weaken the impact from physical boundary.

In this paper,  the scheme of proper time regularization is adopted  [28].
   The virtue  of using proper time regularization here is that it enables us  to perform the sum
  over the thermal Matsubara frequencies analytically. Other than most previous works which use cutoff scheme,  we first use the proper time regularization to study the pion-pion scattering length in a finite volume size.   We expect to extract the signals of chiral  phase transition
 from  pion-pion interactions in the hot medium  while considering the  finite volume effect.

 The paper is organized as follows. In section \ref{sec.model},  the volume and temperature dependence of effective quark mass is deduced from Nambu--Jona-Lasinio (NJL)  model under the proper time regularization. In section \ref{sec.pion},    the   equations for the   meson mass and the pion decay constant at finite volume are deduced. We also give numerical results and analyze them in detail. The  scattering length formula and its corresponding numerical results are presented in section \ref{sec.sca},  closely followed by a short conclusion  in section \ref{sec.conclude}.

\section{NJL model at finite volume size} \label{sec.model}

We use a two-flavor NJL Lagrangian model,  which is motivated by the symmetries of QCD,   to describe  the coupling between quarks and the chiral condensate in the scalar-pseudoscalar sector. It reads as  [29,30]
\begin{eqnarray}\label{eq.L}
	\mathcal{L}_{NJL} & =&\bar \psi (i\gamma_\mu \partial^\mu-  m_q)\psi  \nonumber \\
	&+&G[(\bar \psi  \psi)^2+(\bar \psi i\gamma_5\tau  \psi)^2],
\end{eqnarray}
where $m_q$ is the current quark mass of flavor $q$ and $G$ is the four quark effective coupling.  In the limit of exact isospin symmetry,  $m_u=m_d=m$. The pion mass  can be exploited  through the effective interaction for the exchange of a pion in the random-phase approximation.

 The second term of four quark interaction in Eq. (\ref{eq.L}) is responsible for exciting the pion  as an isovector pseudoscalar.
 In the mean field approximation,  the effective quark mass is
 \begin{equation}\label{eq.s1}
 M=m+\sigma
 \end{equation}
  with
 \begin{equation}\label{eq.s2}
 \sigma=-2G\left\langle {  \bar \psi \psi} \right\rangle
 \end{equation}
 and
   the two-quark condensate is defined as
\begin{equation}\label{eq.gap1}
\left\langle {  \bar \psi \psi} \right\rangle=-\int \frac{d^4p}{(2\pi)^4}Tr[S(p)],
\end{equation}
where $S(p)$ is the dressed quark propagator and the trace is taken in color,  flavor and Dirac spaces.

The integration of Eq.(\ref{eq.gap1}) is ultraviolet divergent.  For simplicity a cutoff on the  momentum  integration is usually taken which  is valid when the cutoff is lager than the relevant momenta.   We will use   the proper time regularization [31-36] for two reasons. Firstly,  we will work with a period or anti-period  boundary condition but the  cutoff on the  momentum breaks the symmetry in the spatial direction. Secondly,   a translation of the quark momentum is required in deducing  the equation of meson mass which   requires the cutoff   to tend to infinity. The proper time method could overcome these difficulties by introducing a new integration.
  Under this kind of regularization scheme  the  trace term in Eq.(\ref{eq.gap1}) is replaced by an integral with
 a suitable choice of the cutoff function. Here in the quark gap equation  the key equation is a replacement
  \begin{equation}
  \frac{1}{A^n } \rightarrow \frac{1}{(n-1) !}\int_{\tau_{UV}}^ {\tau_{IR}} d\tau \tau^{n-1}e^{-\tau A }.
  \end{equation}
 Here,  $\tau_{UV}$ is introduced to regularize the ultra-violet divergent and an infrared cutoff $\tau_{IR}$ is   adopted which  appears also in the literature  [36-37].

By using a Wick rotation,  the two quark condensate  at   infinity volume and   zero temperature  can be written as
\begin{eqnarray}\label{eq.gaps}
\left\langle {  \bar \psi \psi} \right\rangle&=&-N_cN_f\int\frac{d^4p}{(2\pi)^4}\frac{4M}{p^2+M^2}  \nonumber \\
&=&-24M\int^\infty_{-\infty} \frac{d^4p}{(2\pi)^4}\int^  {\tau_{IR}}_{\tau_{UV}} d\tau e^{-\tau(p^2+M^2)}  \nonumber\\
&=&-\frac{3M}{2\pi^2}\int^  {\tau_{IR}} _{\tau_{UV}} d\tau\frac{ e^{-\tau M^2}}{\tau^2},
\end{eqnarray}
with the number of color is $N_c=3$ and the number of flavor $N_f=2$.
 At non-zero temperature,   the quark four-momentum is  replaced by $p_k=(\vec{ p},  \omega_k)$,  with $\omega_k=(2k+1)\pi T$,  $k \in  \mathbb{Z}$. The integration on fourth momentum  in Eq. (\ref{eq.gaps}) is replaced by a sum of all the fermion Matsubara frequencies $\omega_k$.
  The two-quark condensate is given by
\begin{eqnarray}
\left\langle {  \bar \psi \psi} \right\rangle&=&-24M \int^  {\tau_{IR}} _{\tau_{UV}} d\tau e^{-\tau M^2} \times \nonumber\\
&&[T\sum^\infty_{k=-\infty}\int^\infty_{0} \frac{dp}{2\pi^2} \vec{p }^2e^{-\tau(   \vec{p} ^2+\omega_k^2)}]   \nonumber  \\
&=&-\frac{3MT}{\pi^{3/2}}\int^   {\tau_{IR}} _{\tau_{UV}} d\tau\frac{ e^{-\tau M^2}}{\tau^{3/2}}\theta_2(0, e^{-4\pi^2\tau T^2}),
\end{eqnarray}
where the Jacobi function is defined as $\theta_2(0, q)=2\sqrt[4]{q}\sum\nolimits_{n=0}^\infty q^{n(n+1)}.$
Then the quark mass  is
\begin{eqnarray}
M=m+\frac{6GMT}{\pi^{3/2}}\int^   {\tau_{IR}} _{\tau_{UV}} d\tau\frac{ e^{-\tau M^2}}{\tau^{3/2}}\theta_2(0, e^{-4\pi^2\tau T^2}).
\end{eqnarray}

For a specific boundary condition of finite volume,  the quark momentum is  discretized and the integral over all spatial momenta is replaced by sum over discrete momentum modes.
Considering a cubic box with volume size $L$,
the discrete momenta that
depend  on the boundary conditions are
\begin{eqnarray}
 \vec{p}_{PBC}^2&=&\frac{4\pi^2}{L^2}\sum\nolimits_{i=1}^{3}n_i^2,  \qquad n_i=0, \pm 1, \pm 2\cdots, \\
\vec{ p}_{APBC}^2&=&\frac{4\pi^2}{L^2}\sum\nolimits_{i=1}^{3}\left(n_i+\frac{1}{2}\right)^2, \quad n_i= \pm 1, \pm 2\cdots ~~~~~~\label{eq.papbc}
 \end{eqnarray}
 for period and anti-period boundary conditions,
 respectively.
The integration measure is replaced by sum of  discrete momentum with replacement of
\begin{equation}
\int  dp \left(\cdots\right)\rightarrow\frac{2\pi}{L}\sum_{ n_i  }\left(  \cdots\right).
\end{equation}
Then the quark mass is constrained by
\begin{eqnarray}\label{eq.gap3}
M&=&m+48GM\int^ {\tau_{IR}} _{\tau_{UV}}  d\tau e^{-\tau M^2}[T \times  \nonumber \\
& &\sum^\infty_{k=-\infty}e^{-\tau \omega_k^2}\prod \limits_{i=1}^3 \sum_{n_i  }  e^{-\tau   p_i^2}]  \nonumber  \\
&=&m+48GMT\int^ {\tau_{IR}}_{\tau_{UV}} d\tau e^{-\tau M^2}\times  \nonumber \\
&&\theta_2(0, e^{-4\pi^2\tau T^2}) \left[\frac{f(\theta)}{L}\right]^3,
\end{eqnarray}
with
\begin{eqnarray}\label{eq.ftheta}
  { f(\theta)  }&=&\left\{ \begin{array}{l}
 {
   \theta_2(0, e^{-4\tau \pi^2/L^2})      \qquad   \qquad \text{for APBC;}}  \\
   {\theta_3(0, e^{-4\tau \pi^2/L^2})    \qquad  \qquad \text{for PBC;}}      \\
 \end{array} \right.
\label{psix}
\end{eqnarray}
Here  $\theta_3(0, q)=1+2\sum\nolimits_{n=1}^\infty q^{n^2}$,  while  $\theta_2$ is defined as before. Although the momentum integrals
are given by two different functions,  it can be checked that these two functions approach a same limit as $L$ increases to very large value. Thus,  the quark mass will not depend on the boundary condition in the infinite volume limit.

\section{Pion mass and decay constant  }
\label{sec.pion}

The pion,  associated with the exact $ SU_L(2)\times SU_R(2) $ symmetry,  occupies a special place in nuclear and particle physics.  It is   the most relevant degree of freedom  in the low energy regime of the strong interaction,  and is both    (pseudo-)Nambu-Goldstone  bosons and
   quark-antiquark bound-states.   The pion mass and decay constant are measures for the strength of  chiral  symmetry breaking   [13,38].

In the Lagrangian of Eq. {\eqref{eq.gap1}},  the four quark interaction   term $(\bar \psi \psi)^2$  is associated with the scalar $\sigma$ meson while  term $(\bar \psi i\gamma_5\tau  \psi)^2$ the $\pi$ meson. By comparing with the amplitude of $qq$ scattering for the exchange of a pion,   the meson mass is deduced from the proper polarization insertion of the four quark interaction.  In the random phase approximation,  the polarization insertion is related to quark propagator with
\begin{eqnarray}\label{eq.polar1}
\Pi_{ps}(k^2)\delta_{\alpha\beta} =\qquad\qquad\qquad\qquad\qquad\qquad\qquad\qquad\nonumber\\
 -i\int\frac{d^4p}{(2\pi)^4}Tr[i\gamma_5T_\alpha iS(p+k/2)i\gamma_5T_\beta iS(p-k/2)].~~~~~
\end{eqnarray}
Here $T_i$ acts on the external quarks,  $p$ is the quark momentum and $k$ is the meson momentum. The trace is taken in
 Dirac,  flavor and color space. Then the meson mass is the solution of
\begin{equation} \label{eq.gpi}
1-2G\Pi_{ps}(k^2=m^2)=0,
\end{equation}
and the effective coupling strength between pion meson and quarks $ g_{\pi qq} $ is defined as
\begin{eqnarray}\label{eq.gpp}
	g_{\pi q q}^{2}=({\partial \Pi_{ps}}/{\partial k^2})^{-1}|_{k^2=m^2}.
\end{eqnarray}

Performing the trace in Eq. (\ref{eq.polar1})   the proper polarization is
\begin{eqnarray}\label{eq.polar}
\Pi_{ps}(k^2) &=&-4iN_cN_f\int\frac{d^4p} {(2\pi)^4} \times\nonumber\\
&& \frac{M^2+p^2-k^2/4}{[(p+k/2)^2-M^2][(p-k/2)^2-M^2]}.~~
\end{eqnarray}
After making appropriate shifts of quark momentum we can get
\begin{eqnarray}\label{eq.polar2}
\dfrac{1}{i}\Pi_{ps}(k^2) &=& 4N_cN_f\int\frac{d^4p} {(2\pi)^4} \frac{1}{p^2-M^2}\nonumber \\
&&-2N_cN_f   k^2I(k)
\end{eqnarray}
with
\begin{eqnarray}
I(k)=\int\frac{d^4p}{(2\pi)^4}\frac{1}{[(p+k)^2-M^2](p^2-M^2)}.
\end{eqnarray}
Comparing Eq. (\ref{eq.gap1}) with  Eq. (\ref{eq.polar2}),   we have
\begin{equation}
2G\Pi_{ps}=1-\frac{m}{M}+4iGN_cN_fm_{\pi}^2 I(m_{\pi}^2).
\end{equation}
By inserting it into  Eq. (\ref{eq.gpi}),  the mass of pion meson   is deduced with
\begin{equation}\label{eq.pmass}
m_\pi^2=-\frac{m}{M}\frac{1}{4iGN_cN_f I(m_{\pi}^2)}.
\end{equation}
The meson mass   is proportional  to the current quark mass and depends on the effective quark mass. In the chiral limit with $m=0$,  the meson mass is $m_\pi^2=0$.

The  main task now is to calculate the integration of $I(k)$. After Wick rotation the calculation on $I(k)$ is available by introducing the
  Feynman parameters which gives
\begin{eqnarray}
I(k)&=&\int_0^1 dz   \int\frac{d^4p}{(2\pi)^4} \times \nonumber\\
&&\frac{1}{  \{[p - k (1 - z)]^2 - M^2 + k^2 (1 - z) z\}^2}.
\end{eqnarray}
In the  proper time scheme,   we have
 \begin{eqnarray}
I(m_\pi^2)&=&\int_0^1 dz  \int_{\tau_{UV}}^{\tau_{IR}} d\tau \int\frac{d^4p}{(2\pi)^4} \times \nonumber\\
&& {  \tau e^{-\tau[ p  ^2 + M^2 -m_\pi^2 (1 - z) z]}}.
\end{eqnarray}
At nonzero temperature and finite volume size,  the discretizing in temporal and spatial directions gives
 \begin{eqnarray}\label{eq.gappi}
I(m_\pi^2)&=&T\int_0^1 dz  \int_{\tau_{UV}}^{\tau_{IR}}  d\tau   e^{-\tau [M^2 -m_\pi^2 (1 - z) z]}\times  \nonumber \\
&&\tau \theta_2(0, e^{-4\pi^2\tau T^2}) \left[\frac{f(\theta)}{L}\right]^3,
\end{eqnarray}
where $f(\theta)$ is defined as in Eq. (\ref{eq.ftheta}).

 Pion decay constant is an important quantity in low energy phenomena which can be extracted from measurements of the decay $\pi^-\to\mu^-+\nu_\mu$.
From the vacuum to one pion and axial vector current matrix element $ \left<0 \left| J^i_{5\mu}\right|\pi^j\right> $,  the decay constant is given by
\begin{eqnarray}\label{eq.fpiv}
ik_\mu f_\pi\delta^{ij}&=&-N_cg_{\pi qq } \int\dfrac{d^4p}{(2\pi)^4}tr[\gamma_\mu\gamma_5
\nonumber\\
& &\times   S(p+\frac{1}{2}k)\gamma_5S(p-\frac{1}{2})k].
\end{eqnarray}
Using the relation$  tr \tau ^i\tau^j=2\delta^{ij} $
 and performing the traces over spin labels yields
\begin{equation}
	ik_\mu f_\mu=N_c g_{\pi qq}4 Mk_\mu I(k^2).
\end{equation}

The effective coupling $ g_{\pi qq} $ is obtained by
substituting Eq.\eqref{eq.polar2} into Eq.\eqref{eq.gpp}. Then the square of decay constant is
\begin{equation}\label{eq.decay1}
	f_\pi^2=-4iN_cN_fM^2\dfrac{I^2(m_\pi)}{I(0)+I(m_\pi)-m_\pi^2K(m_\pi)}.
\end{equation}
The equation for the pion decay constant,  independent of the regularization scheme,  is related to the quark mass. And we can get a relation
\begin{equation}\label{eq.fg}
	f_\pi^2g_{\pi qq}^2=4M^2\dfrac{I^2(m_\pi)}{[I(0)+I(m_\pi)-m_\pi^2K(m_\pi)]^2}.
\end{equation}

The function $ K $ is defined here as
\begin{eqnarray}
	K(k)=\int\frac{d^4p}{(2\pi)^4}\frac{1}{[(p+k)^2-M^2](p^2-M^2)^2},  \label{eq.Kk}
\end{eqnarray}
and  it can be calculated through $ I(k) $ with
\begin{eqnarray}
	K(k)=\dfrac{1}{k^2-4M^2}\left[12M^2
	\dfrac{dI}{dk^2}(0)-I(k)+I(0)\right]. ~~
\end{eqnarray}
 However,  to calculate the function $  K(k) $  and latter the function $ L(k) $ directly from Eq. \eqref{eq.Kk},  we  can  use the  following Feynman parameter formula
 \begin{eqnarray}
 	\dfrac{1}{A^nB^m}&=&\dfrac{(n+m-1)!}{(n-1)!(m-1)!} \nonumber\\
 	& &\times \int_0^1dx\dfrac{x^{n-1}(1-x)^{m-1}}{[Ax+B(1-x)]^{n+m}}. ~~~~~~~
 \end{eqnarray}

For  an approximation,    taking the function $ I(k) $ as a smooth function of $ k^2 $,    we have  $ K(k)=0 $.  Then,  the decay constant is simplified to

\begin{equation}\label{eq.decay}
f_\pi^2=-4iN_cN_fM^2I(0).
\end{equation}
As the meson mass is small,    the combination of Eq.(\ref{eq.pmass}) for the pion mass with Eq.(\ref{eq.decay}) for the decay constant gives
\begin{equation}\label{eq.gmor1}
m_\pi^2f_\pi^2=mM(N_fG)^{-1}.
\end{equation}
Comparing with Eq.(\ref{eq.s1}) and Eq.(\ref{eq.s2}) for a small $m$,   we have
\begin{equation}\label{eq.gmor}
m_\pi^2f_\pi^2=-m\left<\bar\psi\psi \right>,
\end{equation}
which is the lowest order approximation to the current algebra result   and called as   Gell-Mann--Oakes--Renner (GOR) relation [39].

 The parameters we used   in this paper are $m=4.8$ MeV,  $G=3.19*10^{-6}$ MeV$^{-2}$,     $\tau_{UV}=1/1080^{2}$ MeV $^{-2}$,  and $\tau_{IR}=1/190^{2}$ MeV$ ^{-2} $.
In this parameter set,  the  effective quark mass is 202.3 MeV and the pion meson mass is 135.1 MeV. The quark mass scaled to the mass at zero temperature is only slightly dependent on different set of parameters [36].
And we will not have the coupling $G$ depending on the volume size $L$ and the other properties.
The corresponding decay constant $ f_\pi =93.0 $ MeV and from the  Weinberg's  formula  the scalar pion-pion scattering lengths in unit of $ m_\pi^{-1} $ are $  a_0=0.147 $ and $ a_2=-0.042 $,   respectively.

\begin{figure}[htbp]
	\centering
	\includegraphics[width=0.35\textheight]{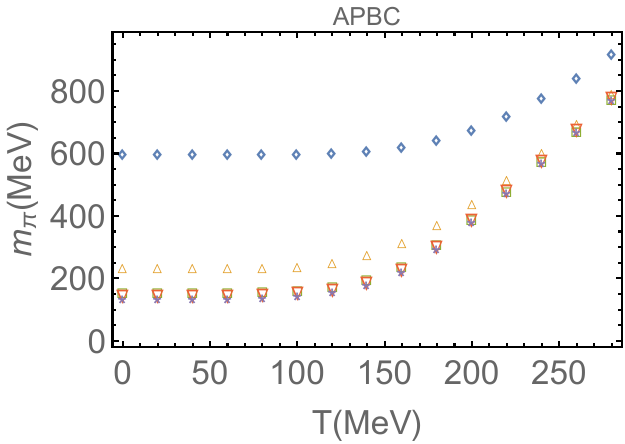}\\
	\includegraphics[width=0.35\textheight]{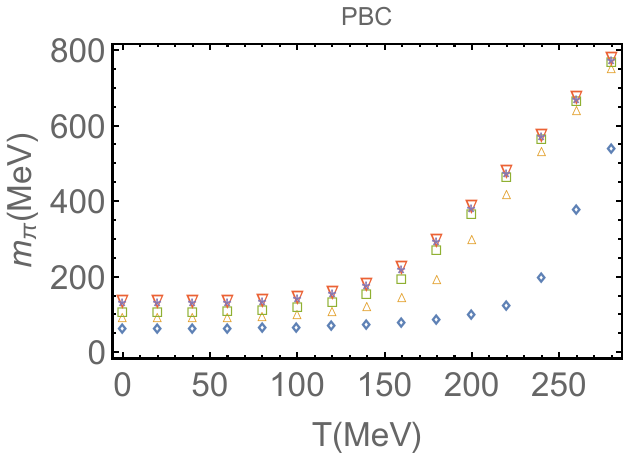}\\
	\caption{Pion  meson  mass as a function of temperature with different boundary condtion and volume size. The plot markers $  \Diamond ,   \Delta , \square ,   \nabla $ and   $ \star $ stand for the volume size $L=1,  1.5,  2,  3$ and  5 fm,  respectively.
	}
	\label{fig.mass}
\end{figure}

\begin{figure}[htbp]
	\centering
	\includegraphics[width=0.35\textheight]{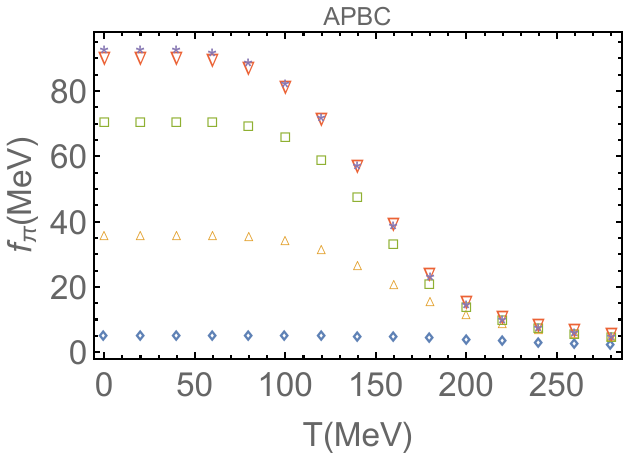}\\ 	\includegraphics[width=0.35\textheight]{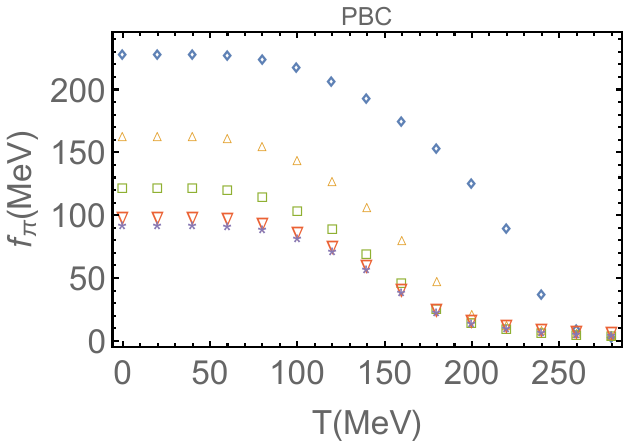}\\
	\caption{ Decay constant as a function of temperature with different boundary  condition  and volume size.  The plot markers $  \Diamond ,   \Delta , \square ,   \nabla $ and   $ \star $ stand for the volume size $L=1,  1.5,  2,  3$ and  5 fm,  respectively.}  \label{fig.decay}
\end{figure}

 The pion mass and decay constant as functions of temperature with different cubic volume size are presented in Fig. \ref{fig.mass} and Fig. \ref{fig.decay},  respectively.  The pion mass increases  and the decay constant decreases as temperature increases.
  The data with volume size $ L $ larger than 5 \text{fm} is close to the infinite volume limit.

	 The GOR relation only keeps well in low temperature and in the infinite volume limit.
	 On the right hand side of the GOR relation Eq. \eqref{eq.gmor},  the current quark mass   comes from  the Higgs mechanism which does not depend on the size effect and is a constant parameter in NJL model here. The chiral symmetry is almost restored at high temperature in finite size,  so the quark condensate is close to zero.  Then  the right hand side of GOR equation is almost temperature and volume size independent. But on the left hand side of the   equation the pion mass times decay constant still depend on T,  which makes the GOR no longer hold.

 Furthermore,  with different choice of boundary conditions,   the influence of volume size,  both on the mass and decay constant,   are different.  It will also reflect on the scattering length.

\begin{figure}[htbp]
	\centering
	\includegraphics[width=0.35\textheight]{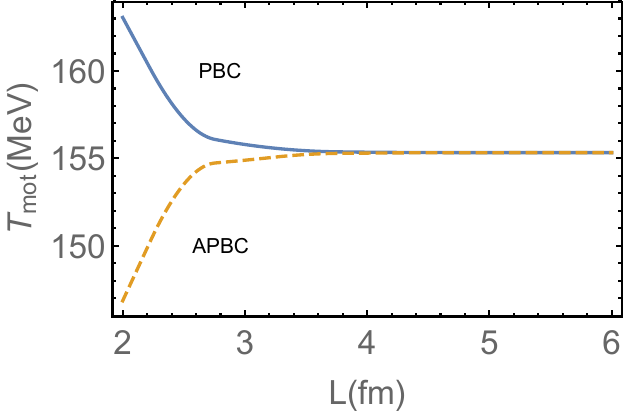}\\
	\caption{ Mott temperature as function of volume size.}  \label{fig.tmot}
\end{figure}

When the temperature increases,  the effective quark mass decreases while the pion mass increases.  As   $ m_q(T) $ = $ 2M(T) $,  which means that the pion can dissociate into a constituent quark and an antiquark,   it defines the Mott temperature $ T_{Mott} $ for pion meson. We can see from Fig. \ref{fig.tmot}  that the Mott temperature is about 155 MeV in the infinity volume limit. In  Fig. \ref{fig.tmot},  we present the Mott temperature as a function of volume size for PBC and APBC. The Mott temperature decreases (increases) with decreasing volume size for APBC(ABC) which may hint that when the volume size effect can not be neglected,   APBC is favored,  at relatively low temperature,  to get the quark gluon plasma.

\section{scattering length}
\label{sec.sca}
From the Weinberg's low energy theorems [40],  the pion scattering lengths   are related to the pion
mass and decay constant,  which also represents a symmetry breaking effect.
The scattering of pion by pions
involves only
the lightest pseudoscalar modes. It provides a direct link between the theoretical formalism of chiral symmetry and experiment.

\begin{figure}[htbp]
	\centering
	\includegraphics[width=0.30\textheight]{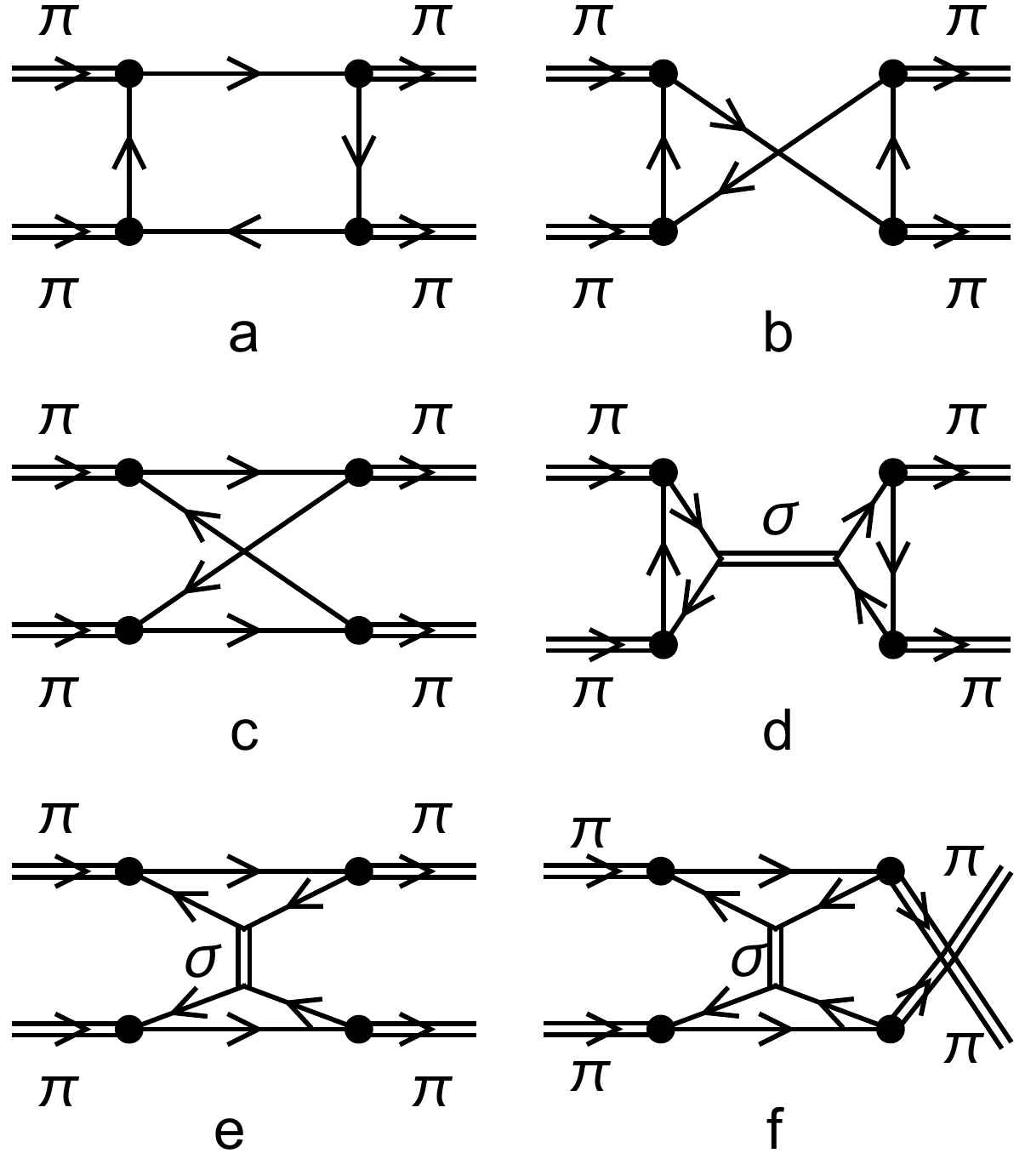}\\

	\caption{ Box and $\sigma$-propagation diagrams for pion-pion scattering.
	}
	\label{fig.feyn}
\end{figure}

Three isospin channels are available for the pion-pion scattering process.
The  invariant  scattering amplitude $\left( \Pi^a\Pi^b\to\Pi^c\Pi^d \right)$ can be  written  as
\begin{eqnarray}\label{eq.Tabc}
	T_{ab, cd}(k, p\to k', p')=A(s, t, u)\delta_{ab}\delta_{cd} \nonumber\\
	+B(s, t, u)\delta_{ac}\delta_{bd}+C(s, t, u)\delta_{ad}\delta_{bc},
\end{eqnarray}
with incoming  momenta $ (k, p, k', p') $ and isospin  indices  $  (a, b, c,   d ) $.
The Mandelstam variables $ s$,  $ t $ and $ u  $ are defined as $  s=(k+p)^2$,  $ t=(k-k')^2 $ and $ u=(k-p')^2 $.
Six  diagrams (box and $ \sigma $-propagation) contribute to pion-pion scattering in the tree level which can be found in Refs. [7-10]. We present then in Fig. \ref{fig.feyn}

The scattering lengths at the kinematic threshold are
\begin{eqnarray}\label{eq.ai}
	a_i=\dfrac{1}{32\pi}A_i(s=4m_\pi^2, t=0, u=0),
\end{eqnarray}
with the three  isospin amplitudes
\begin{eqnarray}\label{eq.abc}
	A_0=3A+B+C, \quad A_1=B-C, \quad A_2=B+C.~~~
\end{eqnarray}
Here,  A,  B and C can be calculated from
$ T_a$,   $ T_b$,   $T_c$,  $T_d$,  $T_e$,  and $T_f$ that correspond to the
amplitudes of the six scattering diagrams.
From those diagrams,  we have $ T_a=T_b $ and $ T_e=T_f $. Then,    A,  B and C in Eq. \eqref{eq.Tabc} and Eq. \eqref{eq.abc} are given by
\begin{eqnarray}
	A=2T_a-T_c+T_d, \quad B=C=T_c+T_d.
\end{eqnarray}
Since $  B=C $,   we have $ A_1=0 $.  So $ a_1 $ can not be calculated in this work. The nonzero amplitudes are
\begin{eqnarray}
	A_0=6T_a-T_c+3T_d+2T_e
\end{eqnarray}
and
\begin{equation}
	A_2=2(T_c+T_e).	
\end{equation}
  Note that,  in most works   a superscript is use to indicate the different isospin.  Since here we only concern the s-wave scattering length and to avoid   confusion with power exponents,  we use a subscript to distinguish the  different isospin  scattering lengths  as in Ref.[9].

 The six $ T_i $ values   corresponding to the scattering diagrams are
\begin{eqnarray}
	T_a  &=& \dfrac{4N}{i}g_{\pi q q}^4[k^2K(k)-I(0)-I(k)],   \\
 	T_c &=& \dfrac{8N}{i}g_{\pi q q}^4 [2k^2K(k)-I(0)-\dfrac{1}{2}k^4L(k)],  \\
 T_ d&=& \dfrac{8N}{i}g_{\pi q q}^4\dfrac{I^2(k)}{(1-k^2/M^2)I(2k)+k^2/(4M^2)I(k)},  ~~~~ \\
 	T_e &=&\dfrac{8N}{i}g_{\pi q q}^4\dfrac{[I(0)-k^2K(k)]^2}{I(0)+k^2/(4M^2)I(k)}, 			
\end{eqnarray}
with $ T_b=T_a $,  $  T_f = T_e$.
Here $  L(k) $  is given by
\begin{eqnarray}
	L(k) =\int\frac{d^4p}{(2\pi)^4}\frac{1}{[(p+k)^2-M^2]^2(p^2-M^2)^2}.
\end{eqnarray}
From Eq.\eqref{eq.fg},  $  g_{\pi qq}^4 $ is given as
\begin{equation}
 g_{\pi qq}^{-4} =-N_c^2N_f^2[I(0)+I(k)-m_\pi^2K(k)]^2.
\end{equation}
with $ k^2=m_\pi^2 $.
After performing the calculations of $  I(k)$,   $L(k) $ and $ K(k) $,  the scattering lengths in Eq.\eqref{eq.ai} can be obtained.

\begin{figure}[htbp]
	\centering
	\includegraphics[width=0.35\textheight]{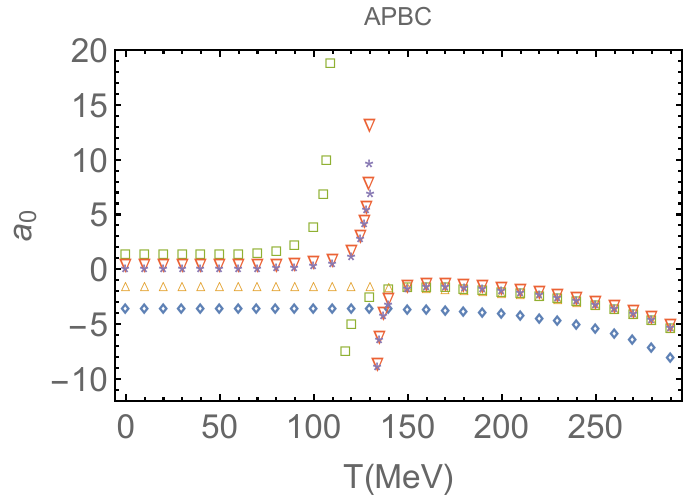}\\ \includegraphics[width=0.35\textheight]{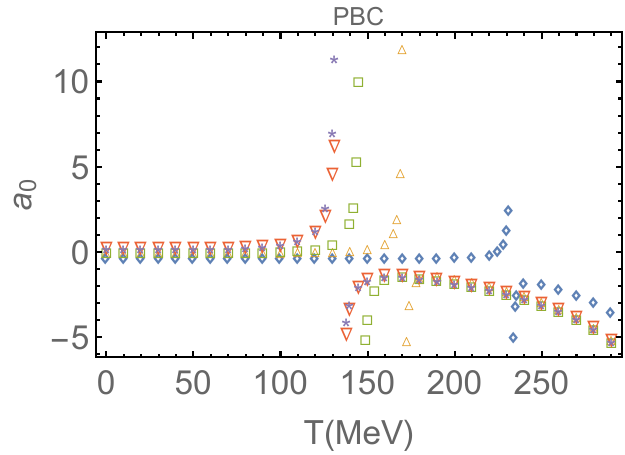}\\
	\caption{The scattering lengths $ a_0 $ for (anti-) period boundary condition.  The plot markers $  \Diamond ,   \Delta , \square ,   \nabla $ and   $ \star $ stand for the volume size $L=1,  1.5,  2,  3$ and  5 fm,  respectively.} \label{fig.a0}
\end{figure}

\begin{figure}[htbp]
	\centering
	\includegraphics[width=0.35\textheight]{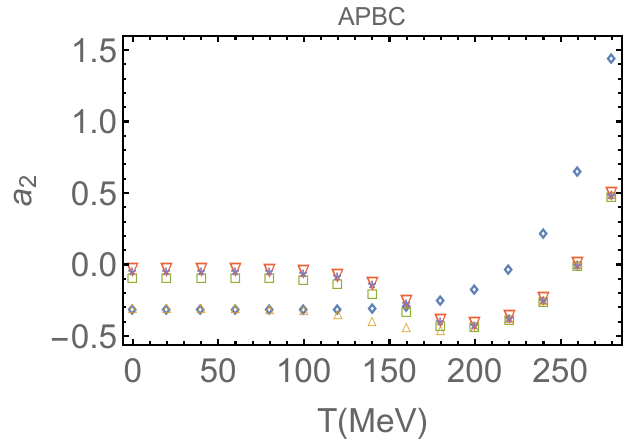}\\ \includegraphics[width=0.35\textheight]{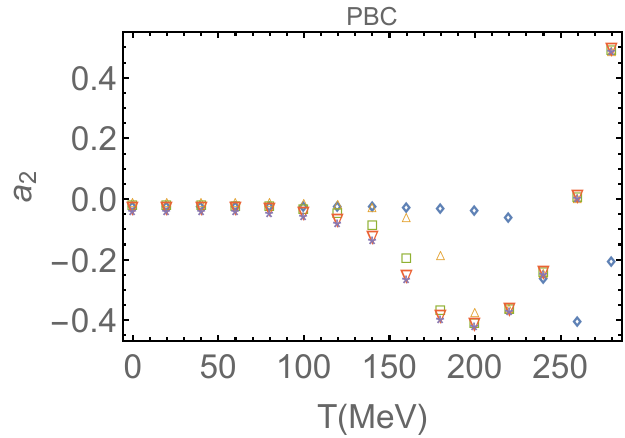}\\
	\caption{The scattering lengths $ a_2 $ for (anti-) period boundary condition.  The plot markers $  \Diamond ,   \Delta , \square ,   \nabla $ and   $ \star $ stand for the volume size 1,  1.5,  2,  3 and  5 fm,  respectively.} \label{fig.a2}
\end{figure}

The numerical results   are presented on Fig. \ref{fig.a0} and Fig. \ref{fig.a2}  for period and antiperiod boundary condition,  respectively.
The scattering lengths are calculated as function of temperature  at several volume sizes. As volume is larger than 5 fm,  the curve is close to the infinite volume limit.  The results  is similar to    results with cutoff regularization that the  $ a_0 $ and $ a_2 $ vary only slowly firstly and then display    steep singularity near the Mott temperature [8,10]. The difference results of the two regularizations appear beyond the Mott temperature. In the cutoff regularization,  because  the chiral  phase transition is first order,  no results beyond the Mott temperature.
 In the proper time regularization,  the chiral phase transition is crossover. So the effective quark mass and scattering lengths can have  continuous values beyond the Mott temperature or pseudo critical temperature of chiral phase transition.

The scattering lengths show apparently   different behaviors   as volume size   decreases for the two kinds of boundary condition. This can be expected from the behaviors of quark mass,  meson mass and decay constant as   the volume size is changing.
Direct  analysis of the Weinberg's formula $ a_0\sim7M_\pi^2/(32\pi f_\pi^2) $ with the results of $M_\pi$ and $f_\pi$ showed in Fig. \ref{fig.mass} and Fig. \ref{fig.decay} gives that  the scattering length  $ a_0 $ at temperature lower than the pseudo critic temperature increases as the volume size decrease for the antiperiod boundary condition,  but  decreases as the volume size decrease for the  period boundary condition.
 The same  analysis  can be used for $ a_2 $ with $ a_2\sim-M_\pi^2/(16\pi f_\pi^2) $.
 However,  we see in Fig.\ref{fig.a0} that,  $  a_0 $ increases and then decreases as volume size decreases at low temperature.

 The scattering length  $ a_0 $ shows jump  as the volume size is not small enough.  We define the jump as the pseudo critical temperature ($ T_c $) which is less than the Mott temperature.  For the period boundary condition,  the jump position in the $ a_0 $ curve increases as volume size decreases and the jump always exists.
 For the antiperiod boundary condition,  $ T_c $ in the $ a_0 $ curve decreases as volume size decreases. As the volume size in small enough the   jump  disappears.
    When the volume size is large enough and  temperature is less than  $ T_c $,  the scattering length $ a_0 $ is larger than zero and  increases with temperature; when temperature is larger than $ T_c $,
 the scattering length $ a_0 $ is less than zero and  increases firstly and then decreases as temperature increases.

  The scattering length $ a_0 $ can be negative at small volume size,  which can be used as an examination of the existence of  phase transition.
Different from the period boundary condition,
 for the antiperiod boundary condition,   $ a_0 $  is a continuous function of temperature at $ L=1 $ fm,   which may due to that chiral symmetry is partly restored at small   finite volume [27].   Since the volume of the smallest QGP system could be as low as (2 fm)$ ^3 $ [18],  a continuous scattering length $ a_0 $ can   serve as a criterion for testing volume effects and different boundary conditions.

 The scattering length $ a_2 $ is always a connected function of temperature. It is also different with the cut off results.
  At very small volume size,  the scattering length $a_2$ is monotonically increases. As the volume size gradually increases,  $a_2$ decreases with the temperature  firstly and then increases. A minimum exists,  but which does not occurred  at $ T_c $.

 For the cut off regularization scheme,  $ a_0 $ increases   within a narrow area near the Mott temperature and both for $ a_0 $ and $ a_2 $   no data exist beyond the Mott temperature  [8,10].

\section{conclusion}
 \label{sec.conclude}.
  In this paper,  we have studied the temperature and volume size dependence of pion decay constant and pion-pion scatterings for different boundary conditions.
  We show the chiral phase transition   of the quark matter at finite temperature and in finite spatial volume.
  Under the proper regularization,  the phase transition indicated from the effective quark mass is crossover   different from the results of  cutoff regularization   where   the phase transition is first order and the   variation of pi-pi scattering length $ a_0 $ with temperature   beyond the Mott temperature between the two regularization scheme.

   The calculated  pi-pi scattering lengths of $ a_0 $ and $ a_2 $   in finite spatial volume  shows different behaviors  for different boundary conditions.   Although we can not determine which boundary condition is the best one,  but the  results  deserve to arouse our attention.

   The pion-pion scattering,  as one of the most fundamental hadronic processes of QCD at the mesonic level,    may serve as a tools to check the different boundary condition and regularization scheme and we hope the results obtained   here may be checked by other theoretical methods.

\section{Acknowledgment}
The authors thank  the Chinese Institute of High Energy Physics and Prof. Huang Mei for authorizing the download  of   the doctoral dissertation.

.

[1] J. Adams et al. (STAR), Nucl. Phys. A 757, 102 (2005).

[2] X. Luo and N. Xu, Nucl. Sci. Tech. 28, 112 (2017).

[3] K. Fukushima and T. Hatsuda, Rept. Prog. Phys. 74,
014001 (2011).

[4] E. Annala, T. Gorda, A. Kurkela, J. N¨attil¨a, and
A. Vuorinen, Nature Phys. 16, 907 (2020).

[5] T. Horn and C. D. Roberts, J. Phys. G 43, 073001 (2016).

[6] V. Pascalutsa and M. Vanderhaeghen, Phys. Rev. D 73,
034003 (2006).

[7] H. J. Schulze, J. Phys. G 21, 185 (1995).

[8] E. Quack, P. Zhuang, Y. Kalinovsky, S. P. Klevansky,
and J. Hufner, Phys. Lett. B 348, 1 (1995).

[9] M. Huang, P. Zhuang, and W. Chao, Phys. Lett. B 465,
55 (1999).

[10] W.-J. Fu and Y.-X. Liu, Phys. Rev. D 79, 074011 (2009).

[11] G. Colangelo, J. Gasser, and H. Leutwyler, Phys. Lett.
B 488, 261 (2000).

[12] Z. T. Draper and S. R. Sharpe, Phys. Rev. D 105, 034508
(2022).

[13] J. Eser and J.-P. Blaizot, Phys. Rev. D 105, 074031
(2022).

[14] G. S. Bali, F. Bruckmann, G. Endrodi, Z. Fodor, S. D.
Katz, and A. Schafer, Phys. Rev. D86, 071502 (2012).

[15] B.-k. Sheng, X. Wang, and L. Yu, Phys. Rev. D 105,
034003 (2022).

[16] Q. W. Wang, Z. F. Cui, and H. S. Zong, Phys. Rev. D
94, 096003 (2016).

[17] Z. Fodor and S. D. Katz, JHEP 04, 050 (2004)

[18] L. F. Palhares, E. S. Fraga, and T. Kodama, J. Phys. G
38, 085101 (2011).

[19] J. Luecker, C. S. Fischer, and R. Williams, Phys. Rev. D
81, 094005 (2010).

[20] J. Gasser and H. Leutwyler, Phys. Lett. B 184, 83 (1987).

[21] J. Braun, B. Klein, and H. J. Pirner, Phys. Rev. D 72,
034017 (2005).

[22] Y. Xia, Q. Wang, H. Feng, and H. Zong, Chinese Physics
C 43, 034101 (2019).

[23] L. M. Abreu, M. Gomes, and A. J. da Silva, Phys. Lett.
B 642, 551 (2006), hep-th/0610111.

[24] K. Saha, S. Ghosh, S. Upadhaya, and S. Maity, Phys.
Rev. D 97, 116020 (2018), 1711.10169.

[25] P. Deb, S. Ghosh, J. Prakash, S. K. Das, and R. Varma,
Chin. Phys. C 46, 044102 (2022), 2005.12037.

[26] B. Klein, Physics Reports 707-708, 1 (2017).

[27] Q. Wang, Y. Xia, and H. Zong, Mod. Phys. Lett. A 33,
1850232 (2018).

[28] J. S. Schwinger, Phys. Rev. 82, 664 (1951).

[29] M. Buballa, Phys. Rept. 407, 205 (2005).

[30] S. P. Klevansky, Rev. Mod. Phys. 64, 649 (1992).

[31] J. Braun, B. Klein, and H. Pirner (2005).

[32] J. Braun, B. Klein, H. J. Pirner, and A. H. Rezaeian,
Phys. Rev. D 73, 074010 (2006).

[33] J. Braun, B. Klein, and H. J. Pirner, Phys. Rev. D 72,
034017 (2005).

[34] Z.-F. Cui, J.-L. Zhang, and H.-S. Zong, Sci. Rep. 7, 45937
(2017).

[35] S.-B. Liao, Phys. Rev. D 53, 2020 (1996).

[36] Z.-F. Cui, C. Shi, W.-M. Sun, Y.-L. Wang, and H.-S.
Zong, Eur. Phys. J. C 74, 2782 (2014).

[37] Y. Ninomiya, W. Bentz, and I. C. Cloet, Physical Review
C 91, 025202 (2015).

[38] T. Horn and C. D. Roberts, J. Phys. G 43, 073001 (2016)

[39] M. Gell-Mann, R. J. Oakes, and B. Renner, Phys. Rev.
175, 2195 (1968).

[40] S. Weinberg, Phys. Rev. Lett. 17, 616 (1966).



\end{document}